\documentclass{article}
\begin{document}
\large
\title{Non-Associative Loops for Holger Bech Nielsen}

\maketitle

\begin{center}
\author{Paul H. Frampton$^{(a)}$, Sheldon L. Glashow$^{(b)}$,\\
Thomas W. Kephart$^{(c)}$ and Ryan M. Rohm$^{(a)}$}\\
\normalsize

\bigskip
\bigskip
\bigskip

$^{(a)}$ Department of Physics and Astronomy,
University of North Carolina,\\
Chapel Hill, NC 27599-3255.\\
$^{(b)}$ Department of Physics, Boston University,
Boston, MA
02215.\\
$^{(c)}$ Department of Physics and Astronomy, Vanderbilt
University,\\
Nashville, TN 37235.
\end{center}

\bigskip
\bigskip
\bigskip

\begin{abstract}
Finite groups are of the  greatest importance in science.
Loops are a simple generalization of finite groups: they
share all the group axioms except for the requirement
that the binary operation be associative.
The least loops that are not themselves groups are
those of order five. We offer a brief discussion of these loops
and challenge the reader (especially Holger)
to find  useful applications
for them  in physics.
\end{abstract}

\bigskip
\bigskip

\section{Introduction}

\bigskip

Many physical systems have symmetries, and groups are the natural
mathematical objects to describe those symmetries (finite groups for
discrete symmetries and infinite continuous groups for continuous
symmetries). If the elements of a group act independently, then the
group is
abelian; if not, it is non-abelian and commutativity amongst the group
elements is lost. For discrete groups, this corresponds to an asymmetry
of the group multiplication table about its principal diagonal, {\it i.
e.,} $ab\neq ba$ for all $a$ and $b\in G$. However, group multiplication
is associative by definition,
\begin{equation}
(ab)c=a(bc)\,,
\end{equation}
and the concept of nonassociative operations  \cite{Schafer} has played a limited role
in science. Nevertheless, it has not been totally absent. Its main
point of entry into physics has been through octonions. Also called
octaves or Cayley numbers, they define
the only division algebra aside from
the real, complex and quarternionic numbers.
An early, but seemingly fruitless, application
of non-associativity in physics is
an octonionic version of quantum mechanics formulated
by Jordan, von Neumann, and Wigner\cite{J,JVW,A}. Attempts have been made
to use
octonions in particle physics to describe quark structure and other aspects of
internal structure.  For reviews see \cite{GunaydinGursey, Costa, Lohmus}. There
are also  an eight-dimensional
octonionic instantons \cite{Grossman, Fubini:1985jm} and applications to
superstrings \cite{Duff:1990wu, Harvey:1991eg}.
Here we observe that the minimal
non-associative  structures are not octonions, but objects called loops.
Let us first define them.

\bigskip
\bigskip

\section{Loops}

\bigskip
\bigskip

A {\bf loop} of order $n$
is a set $L$ of $n$ elements
with a binary operation \cite{Bruck} such that for $a$ and
$b$ elements of $L$, the equations
\begin{equation}
ax=b ~~~~~ and ~~~~~ ya=b
\label{binary}
\end{equation}
each has a unique  solution in $L$.  Furthermore, a loop possesses
an identity element $e$ which satisfies:
\begin{equation}
ex=xe=x\;\;\forall \;x\in L\,
\label{identity}
\end{equation}
The conditions Eqs.(\ref{binary}) and (\ref{identity})
imply that the multiplication table is a Latin square \cite{DK, LM,
Heydayat}.
The multiplication table of a finite group is such
a Latin square, which was defined by Euler as a
square matrix with $n^{2}$ entries of $n$ different elements, none
occuring twice in the same row or column.

Any Latin square whose first row and column are identical defines a {\it
loop} whose upper-left entry is the identity element.  It follows that
any Latin square uniquely defines a loop, although different Latin squares
may define isomorphic loops.  This is because a Latin square remains a Latin
square under any permutation of its columns.  Thus, one can rearrange
any Latin square so that one row is identical to one column.  Once this is
done, that row and column label the elements of the loop and their common
element is the identity element.

\bigskip
\bigskip

A system whose multiplication table has non-identical first
row and column is a quasi-group which
is like a loop but which lacks the identity element
of Eq. (\ref{identity}). We do not consider these structures here.

\bigskip
\bigskip

In contrast to a group multiplication table, the binary opearation
defined by a  Latin square need not
be associative. However, all loops corresponding to
Latin squares with
$n\le 4$ satisfy equation (1). They yield the  groups
$I$, $Z_{2}$ and
Z$_{3}$
at orders 1, 2, and 3, and either
$Z_{2}\times Z_{2}$ or  $Z_{4}$ at order 4.

\bigskip

The situation becomes more
interesting at $n=5$, for which there are five distinct loops. One of
these is
the group $Z_5$. The remaining four are non-associative loops.
For  $n = 6$, there are two
groups, $Z_{2}\times Z_{3}$ and $D_{3}$, and  107 non-associative loops.

\bigskip

The number of non-associative loops rises {\it very} rapidly with $n$ and is
known only for small values.
The number of reduced Latin squares (those in the form with identical
first row and first
column as in all the examples below) is known to be 
9,408; 16,942,080; 535,281,401,856;
377,597,570,964,258,816 and 7,580,721,483,160,132,811,489,280 at orders
n = 6; 7; 8; 9 and 10 respectively. For
n=11 the number of reduced Latin squares, and
hence the (smaller) number of non-associative loops which corresponds to the
number of isomorphism classes of Latin squares which contain at least
one reduced Latin square per class, is not yet known (see {\it e.g}
\cite{DK, Heydayat}).

\bigskip

Loops
are known to arise in the geometry of projective planes \cite{Pickert},
in combinatorics,
in knot theory \cite{Bar-Natan} and in non-associative algebras, but
have yet to play a role  in physics.
Thus  we present all the $n=5$ cases and some (not all!)
of the n=6 non-associative loops as a challenge to Holger and
others, who may find them to be interesting and useful for reasons
too subtle to have been revealed  to us.

\vfill
\newpage

We begin by presenting
all of the five $n=5$ multiplication tables (see p. 129 of\cite{DK})
in a form familiar from group theory. Case (1a) is the group $Z_5$
(the fifth roots of unity)
whilst  the other four are inequivalent non-associative n=5 loops.
 Case (1b) is special in that the square of any
element is the identity element.
As we discuss is \S 3, all 5-loops define commutation algebras that
satisfy the Jacobi identity.
\bigskip
\bigskip

\bigskip
\bigskip
\bigskip
\bigskip
\bigskip

$
\begin{array}{llllll}
\times  & \vline ~~ 1 & 2 & 3 & 4 & 5 \\
\hline 1  & \vline ~~ 1 & 2 & 3 & 4 & 5 \\
2 & \vline ~~ 2 &3 & 4 & 5 & 1 \\
3 & \vline ~~ 3 & 4 & 5 & 1 & 2 \\
4 & \vline ~~ 4 & 5 & 1 & 2 & 3 \\
5 & \vline ~~ 5 & 1 & 2 & 3 & 4
\end{array}
$ $\; \;
\begin{array}{llllll}
\times_{1} & \vline ~~ 1 & 2 & 3 & 4 & 5 \\
\hline 1 & \vline ~~ 1 & 2 & 3 & 4 & 5 \\
2 & \vline ~~ 2 & 1 & 4 & 5 & 3 \\
3 & \vline ~~ 3 & 5 & 1 & 2 & 4 \\
4 & \vline ~~ 4 & 3 & 5 & 1 & 2 \\
5 & \vline ~~ 5 & 4 & 2 & 3 & 1
\end{array}
$ $\;\;
\begin{array}{llllll}
\times_{2} & \vline ~~ 1 & 2 & 3 & 4 & 5 \\
\hline 1 & \vline ~~ 1 & 2 & 3 & 4 & 5 \\
2 & \vline ~~ 2 & 1 & 5 & 3 & 4 \\
3 & \vline ~~ 3 & 4 & 2 & 5 & 1 \\
4 & \vline ~~ 4 & 5 & 1 & 2 & 3 \\
5 & \vline ~~ 5 & 3 & 4 & 1 & 2
\end{array}
$ $\;\;$

(1a)\ \ \ \ \ \ \ \ \ \ \ \ \ \ \ \ \ \ \ \ \ \ \ \ \ \ \ (1b)\ \ \ \ \
\ \
\ \ \ \ \ \ \ \ \ \ \ \ \ \ \ \ \ \ \ \ \ \ \ (1c)

\bigskip
\bigskip
\bigskip
\bigskip
\bigskip
\bigskip
\bigskip
\bigskip
\bigskip
\bigskip

$
\begin{array}{llllll}
\times _{3} & \vline ~~ 1 & 2 & 3 & 4 & 5 \\
\hline
1 & \vline ~~ 1 & 2 & 3 & 4 & 5 \\
2 & \vline ~~ 2 & 1 & 4 & 5 & 3 \\
3 & \vline ~~ 3 & 4 & 5 & 1 & 2 \\
4 & \vline ~~ 4 & 5 & 2 & 3 & 1 \\
5 & \vline ~~ 5 & 3 & 1 & 2 & 4
\end{array}
$ $\;\;
\begin{array}{llllll}
\times _{4} & \vline ~~ 1 & 2 & 3 & 4 & 5 \\
\hline
1 & \vline ~~ 1 & 2 & 3 & 4 & 5 \\
2 & \vline ~~ 2 & 3 & 4 & 5 & 1 \\
3 & \vline ~~ 3 & 5 & 2 & 1 & 4 \\
4 & \vline ~~ 4 & 1 & 5 & 3 & 2 \\
5 & \vline ~~ 5 & 4 & 1 &
2 & 3
\end{array}
$ $\;\;$

(1d)\ \ \ \ \ \ \ \ \ \ \ \ \ \ \ \ \ \ \ \ \ \ \ \ \ \ \ \ \ \ (1e)

\vfill

\newpage

\bigskip
\bigskip

\noindent  Here we present three illustrative examples of the 107
distinct non-associative 6-loops. Each of these defines a commutation
algebra that satisfies the Jacobi identity:

\bigskip
\bigskip
\bigskip
\bigskip

$
\begin{array}{lllllll}
\times _{1}^{6} & \vline ~~ 1 & 2 & 3 & 4 & 5 & 6 \\
\hline
1 & \vline ~~ 1 & 2 & 3 & 4 & 5 & 6 \\
2 & \vline ~~ 2 & 1 & 4 & 3 & 6 & 5 \\
3 & \vline ~~ 3 & 5 & 1 & 6 & 4 & 2 \\
4 & \vline ~~ 4 & 6 & 5 & 1 & 2 & 3 \\
5 & \vline ~~ 5 & 3 & 6 & 2 & 1 & 4 \\
6 & \vline ~~ 6 & 4 & 2 & 5 & 3 & 1
\end{array}
$ $\;\;$ $ \\
$

\bigskip
\bigskip
\bigskip
\bigskip

$
\begin{array}{lllllll}
\times _{2}^{6} & \vline ~~ 1 & 2 & 3 & 4 & 5 & 6 \\
\hline 1 & \vline ~~ 1 & 2 & 3 & 4 & 5 & 6 \\
2 & \vline ~~ 2 & 1 & 6 & 5 & 3 & 4 \\
3 & \vline ~~ 3 & 6 & 1 & 2 & 4 & 5 \\
4 & \vline ~~ 4 & 5 & 2 & 1 & 6 & 3 \\
5 & \vline ~~ 5 & 3 & 4 & 6 & 1 & 2 \\
6 & \vline ~~ 6 & 4 & 5 & 3 & 2 & 1
\end{array}
$  $\;\;$ $ \\
$

\bigskip
\bigskip
\bigskip
\bigskip

$
\begin{array}{lllllll}
\times _{3}^{6} & \vline ~~ 1 & 2 & 3 & 4 & 5 & 6 \\
\hline
1 & \vline ~~ 1 & 2 & 3 & 4 & 5 & 6 \\
2 & \vline ~~ 2 & 1 & 5 & 6 & 4 & 3 \\
3 & \vline ~~ 3 & 4 & 1 & 5 & 6 & 2 \\
4 & \vline ~~ 4 & 3 & 6 & 1 & 2 & 5 \\
5 & \vline ~~ 5 & 6 & 2 & 3 & 1 & 4 \\
6 & \vline ~~ 6 & 5 & 4 & 2 & 3 & 1
\end{array}
$  $\;\;$ $ \\
$

\newpage

The following two examples of non-associative n=6 loops
define commutator algebras that fail to satisfy the Jacobi identity:

\bigskip
\bigskip
\bigskip
\bigskip

$
\begin{array}{lllllll}
\times _{4}^{6} & \vline ~~ 1 & 2 & 3 & 4 & 5 & 6 \\
\hline
1 & \vline ~~ 1 & 2 & 3 & 4 & 5 & 6 \\
2 & \vline ~~ 2 & 1 & 4 & 5 & 6 & 3 \\
3 & \vline ~~ 3 & 6 & 1 & 2 & 4 & 5 \\
4 & \vline ~~ 4 & 5 & 6 & 1 & 3 & 2 \\
5 & \vline ~~ 5 & 3 & 2 & 6 & 1 & 4 \\
6 & \vline ~~ 6 & 4 & 5 & 3 & 2 & 1
\end{array}
$  $\;\;$ $ \\
$

\bigskip
\bigskip
\bigskip
\bigskip

$
\begin{array}{lllllll}
\times _{5}^{6} & \vline 1 & 2 & 3 & 4 & 5 & 6 \\
\hline
1 & \vline 1 & 2 & 3 & 4 & 5 & 6 \\
2 & \vline 2 & 6 & 5 & 1 & 3 & 4 \\
3 & \vline 3 & 1 & 4 & 2 & 6 & 5 \\
4 & \vline 4 & 3 & 6 & 5 & 1 & 2 \\
5 & \vline 5 & 4 & 1 & 6 & 2 & 3 \\
6 & \vline 6 & 5 & 2 & 3 & 4 & 1
\end{array}
$  $\;\;$ $ \\
$

\newpage

\bigskip
\bigskip

\section{Physics Challenge}

\bigskip
\bigskip

In this section  we  suggest a few possible applications of loops
to physics. We challenge the reader
to develop  a useful application to physics from these notions or any
others.
First, it may be useful to point out that the condition of associativity
which is required of groups is a natural condition for symmetry
transformations, since it is an automatic consequence of the composition
of mappings.  Such mappings between particle states, or between states
in a Hilbert space, give rise to the familiar symmetry groups.  Groups
themselves act as transformation groups on themselves, and this action
is consistent with the group action because of associativity.  For a
finite group, for instance, the multiplication table of the group gives a
representation of the group as a set of n permutations
\begin{equation}
g_i(g_j) = \pi_i(g_j) = g_i \times g_j
\end{equation}
and clearly
\begin{equation}
(g_i \times g_j) (g) = g_i (g_j (g) )
\end{equation}
is a consequence of associativity.  For a loop multiplication table,
we again get a set of permutations, but the multiplication by
composition of the permutations is not consistent with the loop multiplication, for
the same reason.  Thus our intuition about groups as transformations may
be a hindrance in interpreting loops in physical applications.

\bigskip
\bigskip

I. One could imagine defining a group product in a way similar to the
definition of a $q$-deformed bosonic commutator algebra where a
fermionic
anticommutator piece is added, $i.e$., here we would consider an
associative
group algebra product $a\cdot b$ deformed by a non-associative loop
algebra
piece $a*b$ to generate an algebra with product
\begin{equation}
a\otimes b=(1-\epsilon )a\cdot b+ \epsilon a*b.
\end{equation}
This may be a way of introducing dissipation or decoherence into a
system.

\bigskip
\bigskip

II. We could try to start with a space $S$ and factor out a loop $L$
similar to an orbifold construction, where a finite group is factored
out.
Such an  $S/L$ loopifold could have application in string theory
although its implementation is made non-trivial by the absence of
matrix representations of the loop.

\bigskip
\bigskip

III.  It is also a consequence of nonassociativity that a representation
of
a loop in terms of linear transformations is never faithful; since
matrix multiplication associates, the nonassociativity must be
annihilated
in any map from the loop to operators on a vector space.  In order
to bypass this obstacle, it is useful to construct  an object familiar
to finite group representation theory, a loop (or group) algebra.  We
take formal linear combinations of the elements of the loop (with
coefficients in ${\bf R}$ or ${\bf C}$), with
multiplication carried out termwise according to the loop multiplication
table.   This procedure defines a vector space whose basis elements are
the loop elements and a natural (but non-associative) multiplication
operation between vectors. We denote the non-associative algebra
corresponding to a group $L$ as $A(L)$.

In particular, the loop elements themselves act as linear
transformations
on $A$ via either left- or right-multiplication.
If $L$ is a group, this action admits  the decomposition
of $A$ into subspaces corresponding to
the irreducible representations
of the group.  For non-associative loops,
the situation is less clear because   matrix
multiplication does not follow the loop multiplication.

However, the algebra associated with a loop has another interesting
property.  To any $A(L)$ (associative or not), we may define the bracket
of two elementa $a\,, \in A$ as
\begin{equation}
[a,b] = a \times b - b \times a\;.
\end{equation}
It is evident that this operation yields an element of $A$, and
furthermore
that it is antisymmetric: $[a,b]=-[b,a]$. However, for non-associative
loops it is far from evident that the bracket operation satisfies the Jacobi
identity,
\begin{equation}
[a,[b,c]] + [b,[c,a]] + [c,[a,b]]=0\;.
\label{jacobi}
\end{equation}
Eq. (\ref{jacobi})  is always satisfied if $L$ is a group.  Every finite
group, through the commutator algebra thus defined, corresponds uniquely to
a Lie algebra.
\medskip
What we find fascinating is that some (but not all) non-associative
loops do yield bracket operations that  satisfy the Jacobi identity,
thereby defining
commutator algebras that are Lie algebras.
Curiously (and as indicated above),
 all of the non-associative loops with $n=5$
are of this class, but only  some of those with n=6.

\bigskip

One could imagine using loops as objects to replace flavor
or horizontal symmetries in particle physics, or using them as
``pregroups."
For example, let us rewrite the $\times _{1}$ loop of Table (1b) in the
form
\bigskip

\bigskip

$
\begin{array}{llllll}
\times _{1} & 1 & a & b & c & d \\
1 & 1 & a & b & c & d \\
a & a & 1 & c & d & b \\
b & b & d & 1 & a & c \\
c & c & b & d & 1 & a \\
d & d & c & a & b & 1
\end{array}
$
\bigskip
\bigskip

\noindent For this case, the bracket operation of the loop
 algebra satisfies the Jacobi identity.
The structure  of the algebra is revealed  in terms of the
linear combinations

\[
K=(a+b+c+d)/2
\]

\[
u_1=(a+b-c-d)/2
\]

\[
u_2=(a-b+c-d)/2
\]

and

\[
u_3=(a-b-c+d)/2
\]
The bracket operation reveals that $K$ (and the identity element)
commute with the other
operations and the $u_i$ \ satisfy the $su_2 $\ algebra
$ [u_i,u_j] = -2\epsilon_{ijk}u_k$.
The nonassociativity lurks
still in the products of these elements, resembling a twisted version of
the Pauli matrices; in this basis they are
given by $K\times u_i=u_i\times K=-u_i/2$, $u_1\times u_2 =3u_3/2$,
$u_2\times u_1 =-u_3/2$, and cyclic permutations of these.
We also have the relations $K^{2}=1+\frac{3K}{2}$ and
$ u_i^{2}=1-K/2$.  It is interesting to note that the combination
$1-K/2$ commutes and associates with the other elements, and the
relation $\Sigma_i u_i^{2}=3(1-K/2) $\ suggests an interpretation
as a Casimir operator for the  $su_2 $; we leave this and other
details for the interested reader to interpret and, hopefully,
apply to physics.

\bigskip
\bigskip

\bigskip
\bigskip

\section*{Acknowledgements}

PHF and RMR acknowledge support by the US Department of Energy under
grant number DE-FG02-97ER-41036. The work of
SLG was supported in part by the National Science Foundation under grant
number NSF-PHY-0099529. TWK was supported by DOE grant number
DE-FG05-85ER-40226.

\end{document}